\documentclass[runningheads]{llncs}
\usepackage[T1]{fontenc}
\usepackage{graphicx}
\usepackage{verbatim}
\usepackage{subcaption}
\usepackage{booktabs}
\usepackage{multirow}
\usepackage{array}
\usepackage[dvipsnames,svgnames,table]{xcolor}
\usepackage{float}
\usepackage{amsmath}
\usepackage{amssymb}
\usepackage{enumitem}
\usepackage{tabularx}
\usepackage{makecell}
\definecolor{darkgreen}{rgb}{0.0,0.5,0.0}
\begin{document}
%
\title{Thinking Like a Scientist: Can Interactive Simulations Foster Critical AI Literacy?} 
\titlerunning{Interactive Simulations for Critical AI Literacy}
%
\author{Yiling Zhao\inst{1}\orcidID{0000-0002-1374-4952} \and
Audrey Michal\inst{2}\orcidID{0000-0002-0304-4012} \and
Nithum Thain\inst{3}\orcidID{0000-0002-7367-0916} \and
Hari Subramonyam\inst{1}\orcidID{0000-0002-3450-0447}}
\authorrunning{Y. Zhao et al.}
%

\institute{Stanford University, Stanford, CA, USA \and
University of Michigan, Flint, MI, USA \and
Google Research, Toronto, Ontario, Canada \\
\email{\{ylzhao,harihars\}@stanford.edu, almichal@umich.edu, nthain@google.com}}
%


\maketitle              
\begin{abstract}
As AI systems shape individual and societal decisions, fostering critical AI literacy is essential. Traditional approaches---such as blog articles, static lessons, and social media discussions---often fail to support deep conceptual understanding and critical engagement. This study examines whether interactive simulations can help learners ``think like a scientist'' by engaging them in hypothesis testing, experimentation, and direct observation of AI behavior. In a controlled study with 605 participants, we assess how interactive AI tutorials impact learning of key concepts such as fairness, dataset representativeness, and bias in language models. Results show that interactive simulations effectively enhance AI literacy across topics, supporting greater knowledge transfer and self-reported confidence, though engagement alone does not predict learning. This work contributes to the growing field of AI literacy education, highlighting how interactive, inquiry-driven methodologies can better equip individuals to critically engage with AI in their daily lives.

\end{abstract}
\section{Introduction}

As artificial intelligence (AI) systems increasingly shape individual and societal experiences, fostering \textit{critical} AI literacy is essential~\cite{popa2024critical}. This literacy extends beyond a conceptual understanding of AI and necessitates critical thinking skills to assess its limitations and broader implications. For example, a hiring algorithm trained on historical recruitment data may inadvertently perpetuate biases, disadvantaging historically underrepresented groups~\cite{raub2018bots}. Thus, it is crucial to recognize that datasets are not neutral --- they are shaped by the choices, assumptions, and limitations of their creators --- and that educators and policymakers must critically analyze training data and decision-making processes to identify and mitigate biases before implementation. 

Current AI literacy approaches employ diverse content and methods, including tools like Google Teachable Machine~\cite{carney2020teachable}, structured curricula integrating technical and ethical topics~\cite{casal2023ai}, hands-on workshops with interactive activities~\cite{hitron2019can,wan2020smileycluster}, and programming-based frameworks such as block-based environments~\cite{jatzlau2019s}. These initiatives are often tailored to specific age groups or audiences, offering accessible introductions to AI concepts while fostering engagement through simplified, project-based learning. However, many of these approaches prioritize \textit{operational skills}, often overlooking the critical thinking required to \textit{interrogate} biases, evaluate limitations, and understand the societal impacts of AI systems. By emphasizing technical knowledge or modular learning, they may neglect the critical literacy needed to navigate AI's ethical, political, and social complexities in real-world contexts~\cite{newman2025ai,de2024human}.

\begin{figure}[h]
    \centering
    \includegraphics[width=1.05\textwidth]{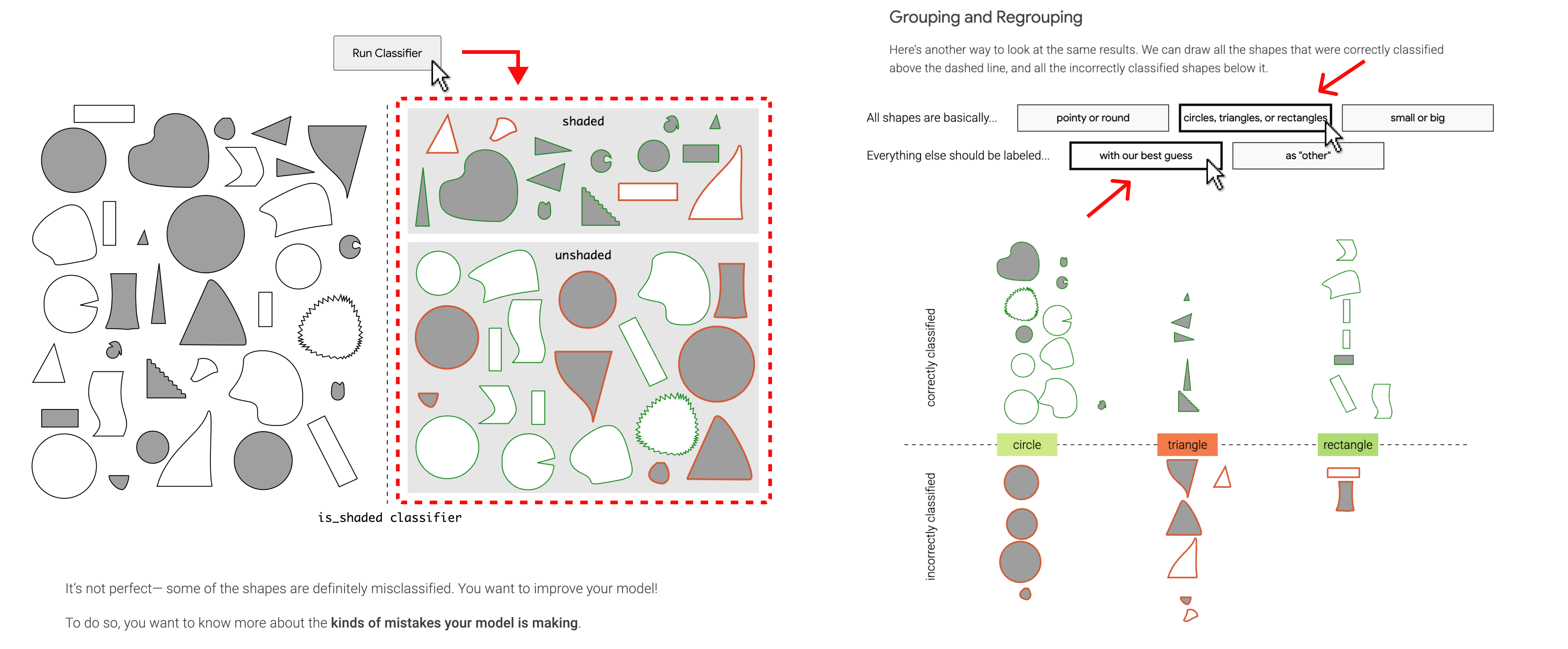}
        \caption{Explorable Illustration: Datasets Have Worldviews}
    \label{fig:explorable-illustration}
\end{figure}

To address this gap, the approach of \textit{scientific discovery learning} (SDL) offers a promising alternative by emphasizing active exploration, hypothesis testing, and iterative reasoning~\cite{de1998scientific}. SDL encourages users to ``think like a scientist'' --- to question assumptions, analyze evidence, and draw conclusions --- potentially enhancing critical thinking and reasoning skills. For instance, an interactive Explorable such as Google's Datasets Have Worldviews~\cite{googlePAIRExplorables} allows people to adjust the parameters of a dataset and observe how differing assumptions, such as changes in category definitions or sampling biases, impact AI classifications (see Figure~\ref{fig:explorable-illustration}). This interactive process could help people understand how the worldviews embedded in datasets shape AI decisions, fostering a deeper understanding of the system's limitations and encouraging critical thinking about the implications of those decisions.

Building on the theoretical foundation of SDL, our work investigates the \textbf{potential of Explorable Explanations to enhance AI literacy by fostering critical thinking and reasoning skills}. Our research questions are: 
\vspace{-5pt}
\begin{description}
    \item[RQ1:] Are interactive tutorial articles effective in teaching AI-related concepts and improving AI literacy?
    \item[RQ2:] If an interactive article effectively improves conceptual understanding of a specific AI ethics topic, does this improvement generalize more broadly across other AI topics?
    \item[RQ3:] What interaction patterns do readers use when interacting with Explorable articles?
\end{description}
\vspace{-5pt}

Through a controlled study involving over 600 participants on Prolific~\cite{prolific}, we evaluate the effectiveness of four Explorable explanations --- focused on datasets and biases, algorithmic decision-making, AI ethics and fairness, and generative AI --- compared to traditional static materials for developing a critical understanding of AI systems. Participants manipulated parameters, tested assumptions, and observed real-time outcomes, aligning with the principles of SDL. Results demonstrated that participants in the Explorable condition experienced significant gains in AI literacy, both in objective assessments and self-reported understanding. While learning gains were also observed in the control conditions, participants in the Explorable condition demonstrated patterns of conceptual improvement and broader generalization across topics, suggesting potential advantages of interactive formats. Furthermore, analysis of interaction patterns revealed that engagement styles varied based on content complexity and topic familiarity. The correlation between interaction quantity and literacy improvement was weak or non-significant, indicating that the quality of engagement, rather than the amount, is crucial. These findings demonstrate the potential of Explorable explanations as an interactive method to address the limitations of current AI literacy approaches. They provide a way to equip end users of AI with the critical thinking and reasoning skills needed to understand and assess AI systems they may encounter.

\vspace{-10pt}
\vspace{-10pt}

\section{Related Work}

AI literacy is a set of competencies that enables individuals to critically evaluate AI technologies, communicate and collaborate effectively with AI, and use AI as a tool online, at home, and in the workplace \cite{long2020ai}. Researchers argue that AI literacy is essential for career development and everyday usage, emphasizing the importance of learning to use AI appropriately while distinguishing between ethical and unethical practices \cite{ng2021conceptualizing}. Cultivating AI literacy equips people with the ability to use AI-driven technologies for advanced learning and skill development \cite{griffin2014assessment}. Furthermore, educating people about AI ethics is critical to promoting the application of AI for societal good \cite{dignum2019responsible}. Hidalgo emphasized that AI literacy should thus encompass both technological aspects of how AI works and its impact on people, including ethics and fairness \cite{hidalgo2024agency}. Recent studies have begun to explore whether training in AI ethics can improve understanding of AI implications. For example, Kasinidou et al. found that attending a brief seminar on Fairness, Accountability, Transparency, and Ethics (FATE) in algorithmic decision-making improved understanding of FATE topics in AI among computer science students \cite{kasinidou2021educating}. Additionally, Kasperson et al. found that having high school students use a web-based tool to learn about machine learning in societal contexts encouraged an improved understanding of AI implications~\cite{kaspersen2022high}.

In addition to investigating \textit{what} content to teach for comprehensive AI literacy, researchers have begun testing \textit{how} to best teach AI literacy. Kim et al. demonstrated that AI education with flipped learning improved academic achievement in students with and without technical backgrounds \cite{kim2023examining}. Alam proposed gaming as a method to assist in teaching AI and machine learning, leveraging its engaging features \cite{alam2022digital}. Others have focused on understanding training data to explain how biases in training data can negatively affect the fairness of AI decisions (i.e., Explainable AI) \cite{anik2021data}. However, Sneirson et al. noted that while several AI practitioners have developed AI explanations (Explainables), they struggle to evaluate the success of their artifacts, and rigorous quantitative pre- and post-test comparisons have been lacking \cite{sneirson2024learning}.

Interactive simulations have shown promising contributions to science education, serving as a platform for inquiry-based learning and fostering the development of scientific knowledge and skills \cite{shved2024teaching}. By supporting sophisticated manipulation of information, interactive tools are useful for teaching complex concepts and simulations \cite{mioduser2000web}. Interactive formats are both engaging and educationally effective across various disciplines, including medical education \cite{hallgren2002interactive,parker2001interactive,servais2006teaching} and engineering \cite{goeser2018view}. However, few studies have specifically investigated the effectiveness of web-based interactive tutorials for teaching the ethical implications of AI. Thus, interactive articles could serve as an entry point for individuals of all ages and backgrounds to build foundational knowledge in AI. Given their potential to explain complex concepts and engage audiences, interactive articles could be particularly effective in improving AI literacy. 

\section{Method}
We conducted a controlled study to evaluate how interactive simulations impact AI literacy. Participants engaged with four AI topics across three different instructional formats, and their learning outcomes were assessed using pre- and post-test measures.

\textbf{Study Design:} Participants were randomly assigned to one of three conditions designed to assess the impact of interactivity on AI literacy: (1) \textit{Explorable (interactive tutorial)}, where participants engaged with a web-based tutorial featuring dynamic visualizations, adjustable parameters, and real-time feedback to explore AI concepts through hands-on experimentation; (2) \textit{Static (PDF tutorial)}, where participants received a non-interactive, text-based version of the tutorial in PDF format, preserving the explanatory content but removing interactive elements such as simulations and input-based feedback; and (3) \textit{Basic Control (no tutorial)}, where participants received no instructional material and proceeded directly from the pre-test to the post-test, serving as a baseline for evaluating prior knowledge and incidental learning. The Explorable and Static conditions allowed for a direct comparison of interactive versus static learning materials, while the Basic Control condition provided a measure of how much learning could be attributed to the interventions rather than general exposure to AI-related concepts.

The \textit{pre-test} included demographic questions (age, gender, ethnicity, education level, and country of residence) and 18 AI literacy items adapted from the Meta AI Literacy Scale~\cite{carolus2023mails}, rated on a 5-point Likert scale from Strongly Disagree (1) to Strongly Agree (5). These questions covered \textit{Use \& Apply AI} (e.g., ``I can operate AI applications in everyday life''), \textit{Know \& Understand AI} (e.g., ``I can assess the limitations and opportunities of using an AI''), \textit{Detect AI} (e.g., ``I can distinguish devices that use AI from devices that do not''), and \textit{AI Ethics} (e.g., ``I can analyze AI-based applications for their ethical implications''). Participants also answered two multiple-choice questions based on real-world AI scenarios from the \textit{AI Incident Database}\footnote{https://incidentdatabase.ai/}, assessing their ability to identify risks and propose mitigation strategies corresponding to the AI topic group they were assigned to. For instance, for the Diversity topic, the case of Incident 37\footnote{https://incidentdatabase.ai/cite/37}: ``Female Applicants Down-Ranked by Amazon Recruiting Tool'' was used to develop a scenario describing how companies use AI to identify the most promising candidates in the resume screening process. The first question asked participants to identify the main potential issue caused by the use of AI in the given scenario, while the second question asked participants to select appropriate mitigation strategies for improving equitable usage.

Following the \textit{pre-test}, participants in the \textit{Explorable} group interacted with an assigned tutorial, while the \textit{Static} group read a non-interactive PDF version. The \textit{Basic Control} group received no instructional material. The \textit{post-test} included additional target-topic AI scenario questions, three non-target AI scenario questions (in randomized order), and a re-evaluation of AI literacy using the same 18 AI literacy items. To analyze engagement for the \textit{Explorable} group, all interactions on the simulation, such as scrolling and clicking, were logged using the participant's session ID along with timestamp information.

\textbf{Topic Selection and Justification:} This study evaluated the effectiveness of interactive simulations in improving AI literacy using four AI Explorables\footnote{https://pair.withgoogle.com/explorables/} developed by a multidisciplinary team at a major technology company. These Explorables were selected to represent diverse AI topics, balancing technical and ethical dimensions: \textbf{Large Language Models}\footnote{https://pair.withgoogle.com/explorables/fill-in-the-blank/}, showcasing advances in language technology; \textbf{Diversity}\footnote{https://pair.withgoogle.com/explorables/measuring-diversity/} and \textbf{Fairness}\footnote{https://pair.withgoogle.com/explorables/measuring-fairness/}, addressing bias and algorithmic equity; and \textbf{Data Representations}\footnote{https://pair.withgoogle.com/explorables/dataset-worldviews/}, illustrating how datasets shape AI outputs. This selection provided a well-rounded foundation for AI literacy, equipping learners with conceptual understanding and critical perspectives on AI’s societal implications.

\textbf{Participants and Recruitment:}  This study was preregistered\footnote{Preregistration available at \url{https://doi.org/10.17605/OSF.IO/F86GA}}, and participants were recruited using the Prolific Platform~\cite{prolific}. They were required to be at least 18 years old, proficient in English, and located in the United States. These criteria helped ensure a broad adult population while controlling for cultural and linguistic variables ($n=612$). Participants in the Explorable and static tutorial control groups were compensated \$4 for completing the survey, while those in the basic control groups were compensated \$2. The compensation was determined based on the relative time required to complete the study, calculated to reflect approximately \$12 for an hour of effort. Qualified participants were randomly assigned to one AI topic group and could not be selected for any other group.

To ensure data quality, we implemented several measures. Explorable group participants were required to interact with the Explorable for at least 5 minutes, with submissions below 1 minute flagged for review. Control group participants had to enter a password at the end of their static tutorial to proceed. Additionally, all participants faced an attention check question in the post-test survey, with incorrect responses leading to submission flagging. These measures collectively enhanced our confidence in the validity of participant engagement and data quality across both groups.

\subsection{Data Analysis}

The survey responses, activity logs, and Prolific submission data were linked using participant's session IDs. Only approved Prolific submissions were retained for analysis. After cleaning and linking, valid data points varied across groups: 47 for LLM Explorable, 51 each for Fairness, Diversity, and Worldview Explorables; 49 for LLM Static control, 50 for Worldview Static control, 51 each for Fairness and Diversity Static controls. The Basic control group consistently had 51 valid data points across all four AI topics.

\textbf{Tutorial Effectiveness:}
We assessed tutorial effectiveness using both within-subject paired \textit{t}-tests and between-group regression analyses. Paired \textit{t}-tests evaluated whether participants demonstrated significant improvements from pre- to post-test on topic-specific scenario items, with $p \leq 0.05$ indicating a meaningful learning gain. To compare instructional formats while accounting for prior knowledge, we fit an ordinary least squares (OLS) regression model predicting post-test scores from pre-test scores and instructional condition.

\textbf{Generalization:} 
To assess knowledge transfer, we compared post-test accuracy between target and non-target topic AI scenario questions using paired \textit{t}-tests. Effective generalization is indicated by statistically similar performance across target and non-target questions ($p > 0.05$). We also conducted an OLS regression to examine whether instructional condition predicted non-target performance after controlling for target-topic scores.

\textbf{Self-assessed AI Literacy:} 
Changes in self-reported AI literacy and ethics understanding were evaluated using paired \textit{t}-tests comparing pre- and post-intervention scores within-subject for each AI topic across all conditions. In addition, we ran an OLS regression to assess the predictive effect of pre-intervention beliefs and instructional condition on post-intervention ratings.

\textbf{Engagement Analysis:} User engagement was analyzed through a multi-faceted approach. We identified distinct scrolling behaviors by visualizing scroll depth over time for individual users, providing insights into various interaction patterns. Additionally, we mapped user interactions into two categories: ``Identify Issue'' and ``Propose Solutions'' activities. To examine how engagement styles related to learning outcomes, we computed partial correlations between activity counts and score improvements, controlling for time-on-task via OLS regression and correlating the resulting residuals. This isolated the unique contribution of interaction types beyond engagement duration.

\section{Results}
The study included a diverse participant pool. The age distribution was right-skewed, with 65\% participants under 40. Gender leaned slightly male (53\% male, 44\% female, 3\% other), and education levels skewed toward higher degrees, with bachelor's holders most common (39\%). Racially, White/Caucasian participants were the largest group (63\%), followed by Black or African American (15.5\%) and Asian participants (11\%). 

\subsection{Effectiveness of Interactive Simulations}
\vspace{-10pt}
\begin{figure}[h!]
    \centering
    \includegraphics[width=0.6\textwidth]{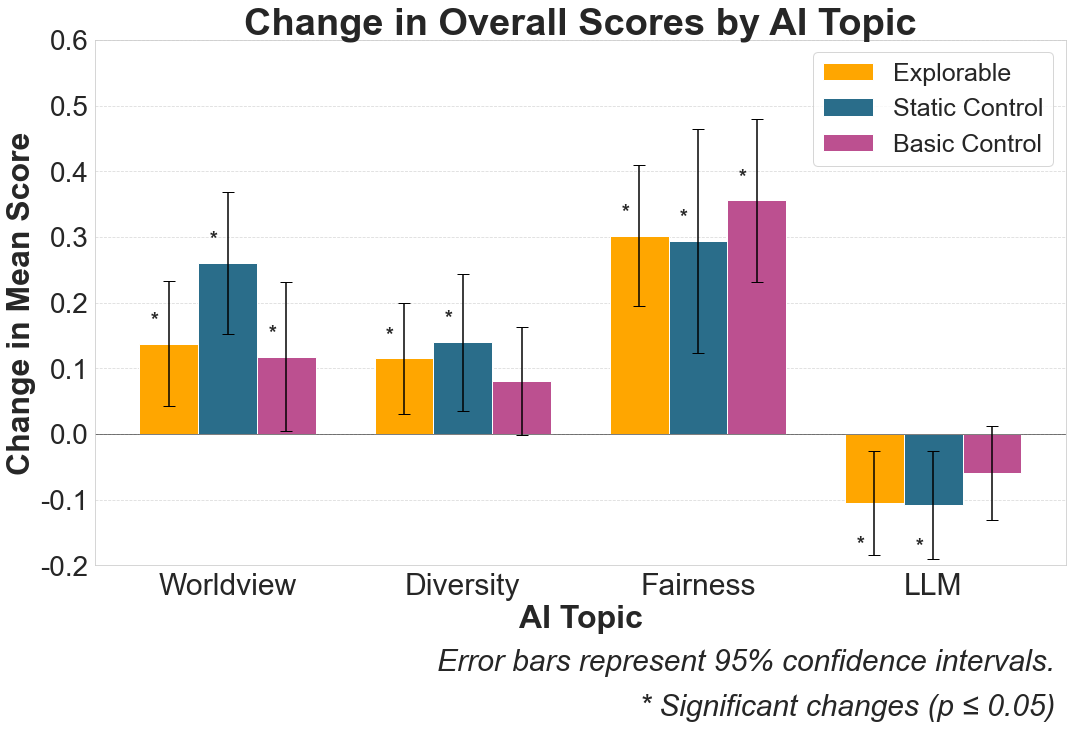}
    \caption{AI Literacy Learning Gains Across Conditions}
    \label{fig:performance_metrics}
\end{figure}
\vspace{-10pt}

Figure~\ref{fig:performance_metrics} displays changes in participants' overall performance on AI literacy questions across four topics: \textit{Worldview}, \textit{Diversity}, \textit{Fairness}, and \textit{LLM}. The Explorable condition showed significant improvements in overall scores across three topics, demonstrating the effectiveness of the interactive intervention. For Worldview, there was a significant gain in ``Identify the issue'' ($p = 0.007$). The Diversity topic showed a significant improvement in ``Propose solutions'' ($p < 0.001$), while Fairness improved consistently across all measures (all $p s \leq 0.008$).

The Explorable group demonstrated statistically significant pre-to-post improvements in most topics, while gains in the Basic Control group were smaller or non-significant. For example, in the Worldview topic, overall scores improved from 0.53 to 0.67 ($p < 0.001$) in the Explorable condition, compared to a smaller increase from 0.46 to 0.58 ($p = 0.047$) in the Basic Control condition. In the Diversity topic, the Explorable group improved significantly from 0.69 to 0.81 ($p < 0.001$), while the Basic Control group’s increase from 0.63 to 0.71 was not statistically significant ($p = 0.06$). Similarly, for ``Propose Solutions'' in Diversity, the Explorable condition improved from 0.65 to 0.86 ($p < 0.001$), while the Basic Control condition showed minimal change (0.69 to 0.72, $p = 0.65$).

Further comparison between the Explorable and Static Control conditions revealed that while both interventions improved AI literacy, their effectiveness varied by topic. The Explorable condition showed clear advantages in fostering solution-oriented thinking for Diversity (0.65 to 0.86, $p < 0.001$; Static Control: 0.62 to 0.74, $p = 0.05$) and Fairness (0.61 to 0.78, $p = 0.01$; Static Control: 0.61 to 0.65, $p = 0.57$). For Worldview, the Static Control was highly effective in improving ``Identify the issue'' scores (0.37 to 0.70, $p < 0.001$), while the Explorable condition also showed significant gains (0.46 to 0.65, $p = 0.01$), suggesting that static materials can also be effective for certain learning outcomes.

Notably, the LLM topic showed mixed results, with all three conditions experiencing some declines in performance. These findings underscore that while interactive and static interventions both enhanced AI literacy in some areas, their relative effectiveness is topic-dependent, highlighting the need to tailor educational strategies to specific AI concepts.

To account for differences in pre-test ability, we conducted a regression analysis predicting post-test scores from pre-test performance and instructional condition:
\[
\text{Post}_{i} = \beta_0 + \beta_1 \cdot \text{Pre}_{i} + \beta_2 \cdot \text{Static}_{i} + \beta_3 \cdot \text{Basic}_{i} + \varepsilon_{i}
\]
\noindent
where \(\text{Static}_{i}\) and \(\text{Basic}_{i}\) are dummy variables and \(\text{Explorable}_{i}\) serves as the reference group.
The model revealed a strong relationship between pre- and post-test performance (\(\beta_1 = 0.22\), SE = 0.03, \(p < .001\)), highlighting the importance of prior knowledge. While post-test scores in the Static (\(\beta_2 = -0.01\), SE = 0.03, \(p = .642\)) and Basic (\(\beta_3 = -0.04\), SE = 0.03, \(p = .127\)) conditions were slightly lower than in the Explorable condition, these differences were not statistically significant. Overall, the regression results support the pattern observed in the within-group analyses: interactive tutorials are at least as effective as static materials and may offer additional benefits for promoting knowledge generalization and building perceptual confidence, as discussed in the following sections.

\subsection{Transfer Task Generalization}
The Explorable condition demonstrated strong generalization across most AI topics. As shown in Table~\ref{tab:combined_results}, there were no significant differences between target and non-target question performance for Worldview ($p = 0.93$), Fairness ($p = 0.16$), and LLM ($p = 0.96$), indicating effective knowledge transfer. However, in the Diversity group, target questions significantly outperformed non-target ones ($p < 0.001$), suggesting that the knowledge acquired in this domain was more specific and less transferable. In contrast, control conditions exhibited limited generalization, with significant target/non-target differences, indicating that interactive learning may better support knowledge transfer than passive exposure.

We also regressed generalization performance (non-target scores) on target performance and instructional condition. The model is specified as:
\[
\text{NonTarget}_{i} = \beta_0 + \beta_1 \cdot \text{Target}_{i} + \beta_2 \cdot \text{Static}_{i} + \beta_3 \cdot \text{Basic}_{i} + \varepsilon_{i}
\]
Target topic performance significantly predicted non-target performance (\(\beta = 0.40, p < .001\)), indicating that learners who understood their assigned topic better also performed better on transfer tasks. Compared to the Explorable condition, the Static control did not differ significantly (\(\beta = -0.06, p = .19\)), while the Basic control scored significantly lower (\(\beta = -0.14, p = .001\)). These results support the hypothesis that interactive simulations better support generalization than basic exposure, even after accounting for topic-specific performance.

\vspace{-20pt}
\begin{table}[htbp]
\centering
\caption*{\scriptsize\textit{Note: ID = Identify the Issue, Prop. Sol. = Propose Solutions, Gen. AI Lit. = General AI Literacy. WV = Worldview, Div = Diversity, Fair = Fairness, Ctrl = Control. Values reflect mean differences (e.g., Target – Non-Target or Post – Pre). Significance: \textsuperscript{*} $p < .05$, \textsuperscript{**} $p < .01$, \textsuperscript{***} $p \leq .001$.}}
\vspace{5pt}
\small
\setlength{\tabcolsep}{5pt}
\begin{tabular}{ll|l l l|l l}
\toprule
 & & \multicolumn{3}{c|}{\textbf{Target vs. Non-Target Items}} & \multicolumn{2}{c}{\textbf{AI Literacy Change}} \\
\textbf{Condition} & \textbf{Topic} & ID Issue & Prop. Sol. & Overall & Gen. AI Lit. & AI Ethics \\
\midrule
\multirow{4}{*}{Explorable} 
& WV   & 0.02 & –0.03 & –0.01 & \textbf{0.11\textsuperscript{*}} & –0.03 \\
& Div  & \textbf{0.18\textsuperscript{***}} & \textbf{0.22\textsuperscript{***}} & \textbf{0.40\textsuperscript{***}} & \textbf{0.16\textsuperscript{***}} & \textbf{0.22\textsuperscript{*}} \\
& Fair & 0.05 & 0.08 & 0.13 & \textbf{0.12\textsuperscript{***}} & 0.12 \\
& LLM  & –0.01 & 0.01 & 0.00 & \textbf{0.21\textsuperscript{***}} & 0.09 \\
\midrule
\multirow{4}{*}{Basic Ctrl} 
& WV   & \textbf{0.17\textsuperscript{***}} & \textbf{–0.14\textsuperscript{*}} & 0.02 & \textbf{0.09\textsuperscript{*}} & 0.11 \\
& Div  & \textbf{0.24\textsuperscript{***}} & \textbf{0.14\textsuperscript{***}} & \textbf{0.38\textsuperscript{***}} & 0.06 & 0.00 \\
& Fair & 0.05 & \textbf{0.18\textsuperscript{***}} & \textbf{0.23\textsuperscript{***}} & 0.03 & 0.05 \\
& LLM  & \textbf{0.14\textsuperscript{*}} & 0.03 & \textbf{0.18\textsuperscript{*}} & 0.08 & 0.12 \\
\midrule
\multirow{4}{*}{Static Ctrl} 
& WV   & \textbf{0.26\textsuperscript{***}} & –0.04 & \textbf{0.22\textsuperscript{*}} & 0.06 & 0.14 \\
& Div  & \textbf{0.16\textsuperscript{*}} & \textbf{0.15\textsuperscript{*}} & \textbf{0.31\textsuperscript{***}} & \textbf{0.10\textsuperscript{*}} & 0.04 \\
& Fair & 0.10 & –0.02 & 0.08 & \textbf{0.10\textsuperscript{*}} & \textbf{0.15\textsuperscript{*}} \\
& LLM  & 0.00 & 0.01 & 0.01 & \textbf{0.15\textsuperscript{***}} & \textbf{0.20\textsuperscript{***}} \\
\bottomrule
\end{tabular}
\caption{AI Literacy Generalization and Self-Assessed Literacy Changes}
\label{tab:combined_results}
\end{table}
\vspace{-20pt}

\subsection{Self-Assessed AI Literacy}
Within-subject paired \textit{t}-tests were conducted to compare pre- and post-intervention self-assessed scores for both general AI literacy and ethical AI literacy. In the Explorable condition, participants reported significant improvements in general AI literacy across all topics. For instance, in the LLM group, the mean self-assessed AI literacy increased from $3.56$ to $3.77$ (\textit{mean difference} = 0.21, $p < 0.001$), and in Diversity, from $3.58$ to $3.74$ (\textit{mean difference} = 0.16, $p = 0.001$). Regarding AI ethics, the Explorable condition showed a significant increase in the Diversity group (\textit{mean difference} = 0.22, $p = 0.03$), while other topics showed more minor or non-significant changes. In contrast, the Basic Control condition showed significant improvement in general AI literacy only for Worldview (\textit{mean difference} = 0.09, $p = 0.01$), and the Static Control condition yielded significant improvements in Diversity ($p = 0.02$), Fairness ($p = 0.01$), and LLM ($p = 0.003$). These findings suggest that interactive experiences enhance participants' confidence in their AI literacy more consistently than those who do not receive AI tutorials.

To further examine between-condition differences in self-assessed AI literacy, we conducted an OLS regression predicting post-intervention scores from pre-intervention beliefs and instructional condition. The model is specified as:
\[
\text{Post\_lit}_{i} = \beta_0 + \beta_1 \cdot \text{Pre\_lit}_{i} + \beta_2 \cdot \text{Static}_{i} + \beta_3 \cdot \text{Basic}_{i} + \varepsilon_{i}
\]
Pre-test scores were a strong predictor of post-test ratings (\(\beta = 0.83, p < .001\)), indicating that initial confidence heavily influenced final beliefs. While participants in the Basic Control condition trended lower than those in the Explorable condition (\(\beta = -0.049, p = .105\)), this difference did not reach statistical significance. No meaningful difference was found between the Explorable and Static Control conditions (\(\beta = -0.0008, p = .978\)). Together, these results reinforce that interactive simulations produce consistent within-subject gains and may offer modest advantages over passive exposure.

\subsection{Interactive Engagement Patterns}
\begin{figure}[htbp]
    \centering
    \begin{subfigure}[b]{0.325\textwidth}
        \centering
        \includegraphics[width=\textwidth]{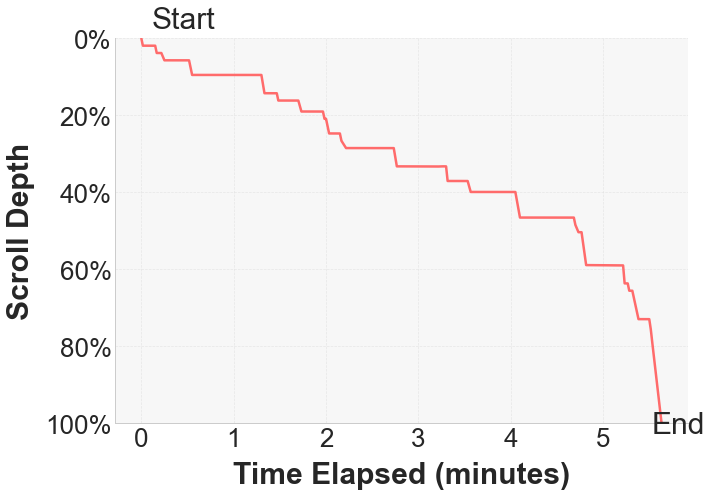}
        \caption{Consistent}
        \label{fig:consistent-scroll}
    \end{subfigure}
    \hfill
    \begin{subfigure}[b]{0.325\textwidth}
        \centering
        \includegraphics[width=\textwidth]{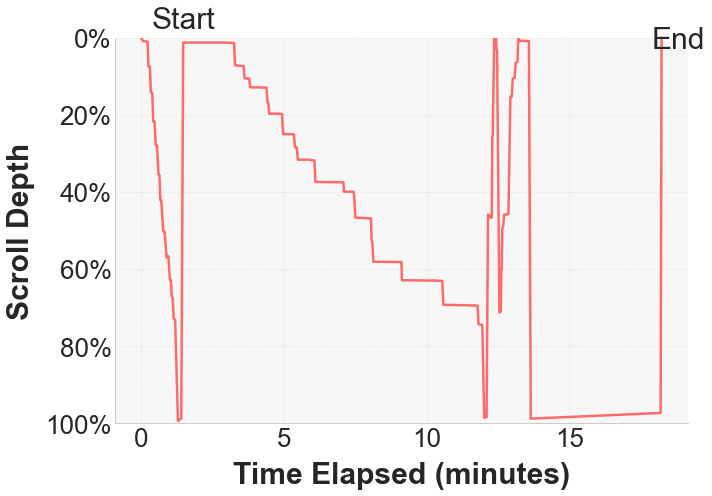}
        \caption{Preview \& Read}
        \label{fig:preview-scroll}
    \end{subfigure}
    \hfill
    \begin{subfigure}[b]{0.325\textwidth}
        \centering
        \includegraphics[width=\textwidth]{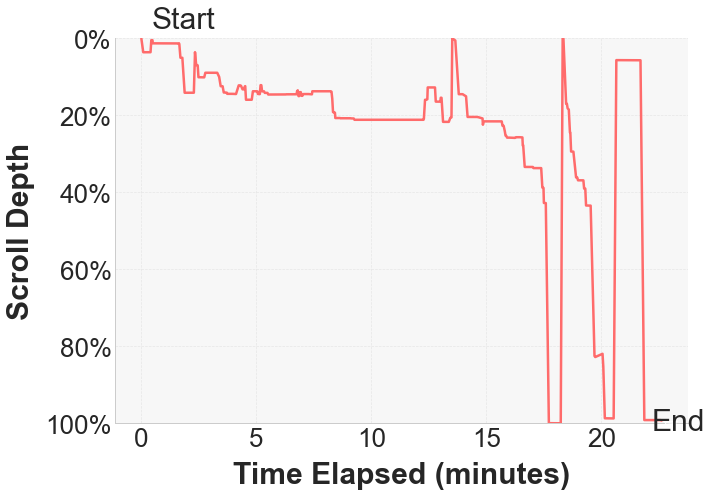}
        \caption{Bounce}
        \label{fig:backtrack-scroll}
    \end{subfigure}
    \caption{Different user scroll patterns}
    \label{fig:scroll-patterns}
\end{figure}

Analysis of user engagement with the Explorable condition revealed diverse scrolling patterns, indicative of different learning approaches. For example, Figure~\ref{fig:consistent-scroll} illustrates a consistent, linear progression, suggesting a methodical and sequential reading style. In contrast, Figure~\ref{fig:preview-scroll} shows an initial rapid scroll—likely to preview content—followed by detailed reading with periods of backtracking and re-reading over 18 minutes, which suggests a more exploratory and reflective approach. Figure~\ref{fig:backtrack-scroll} depicts another pattern, characterized by frequent, short scroll movements over 22 minutes, indicating a careful, section-by-section review and active integration of ideas. These varied engagement styles highlight that learners interact with AI literacy content in multiple ways, which has implications for designing more adaptive and effective educational materials.

\begin{figure}[htbp]
\centering
\includegraphics[width=0.65\textwidth]{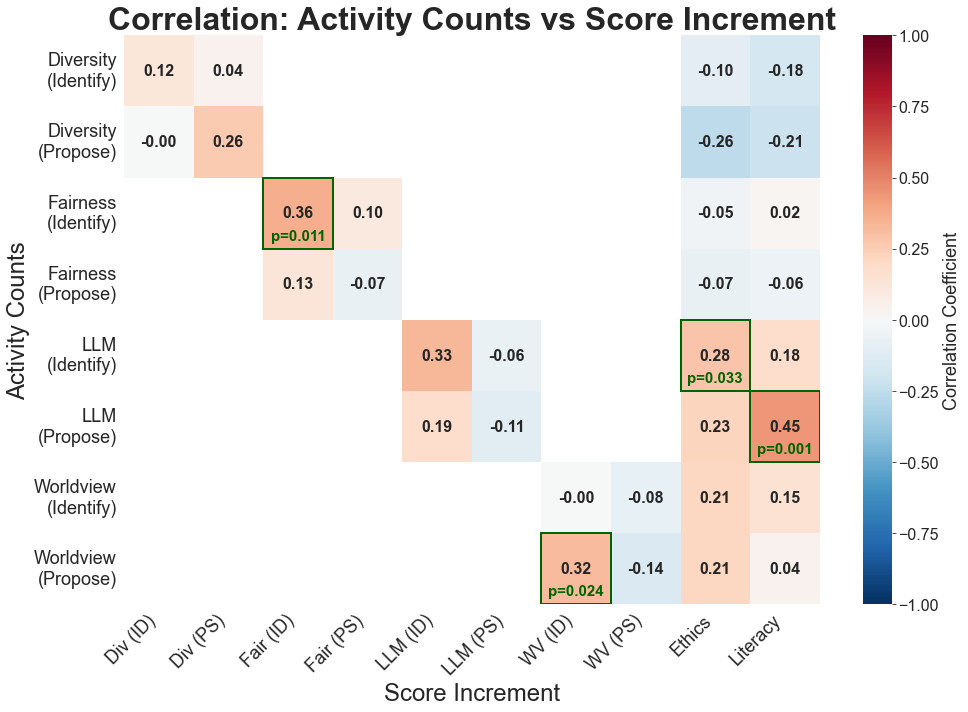}
\caption{Correlation: Activity Counts vs Score Increment}
\label{fig:activity-correlation}
\begin{minipage}{1\textwidth}
\footnotesize
\scriptsize\textit{Note: Abbreviations: ID = Identify Issue, PS = Propose Solution, Div = Diversity, Fair = Fairness, WV = Worldview. Green boxes indicate significant correlations ($p \leq 0.05$).}
\end{minipage}
\end{figure}

User interactions within the Explorable condition were further categorized into ``Identify Issue'' and ``Propose Solutions'' activities. As shown in Figure~\ref{fig:activity-correlation}, partial correlation analyses were conducted between activity counts and score improvements while controlling for time-on-task. Most relationships remained weak or non-significant; however, there were notable exceptions: in the Fairness topic, the number of ``Identify Issue'' actions correlated moderately with performance improvement ($r = 0.36$, $p = 0.011$); in the LLM domain, activity counts correlated with both AI ethics understanding ($r = 0.28$, $p = 0.033$) and overall AI literacy ($r = 0.45$, $p = 0.001$); and in the Worldview topic, ``Propose Solutions'' activities correlated with topic-specific improvements ($r = 0.32$, $p = 0.024$). These findings suggest that while overall engagement does not uniformly predict learning outcomes, the quality and type of engagement in specific tasks is meaningfully related to performance.

\section{Discussion}
\subsection{Implications for AI Education}
Our findings demonstrate that interactive simulations effectively enhance AI literacy, as evidenced by significant within-subject improvements in both objective assessments and self-reported understanding. Participants in the Explorable condition showed notable gains in identifying issues and proposing solutions, particularly within topics such as diversity, fairness, and data representation.

OLS regression analyses indicated that all conditions, including the Static and Basic Controls, yielded some learning gains. Although we did not observe statistically significant differences between conditions, participants in the Explorable condition exhibited stronger patterns of generalization to non-target topics. This suggests that while exposure to AI concepts—such as reading static materials or answering survey questions—can support learning, interactive engagement may facilitate broader conceptual transfer and deeper understanding. The Basic Control group’s limited generalization further indicates that passive exposure may be insufficient for comprehensive learning.

These results align with active learning theories, particularly the ICAP framework \cite{chi2014icap}, which emphasizes that constructive and interactive engagement fosters deeper learning than passive methods. The ability to manipulate parameters, test assumptions, and receive real-time feedback in the Explorable condition likely supported iterative reasoning and reflection, contributing to more durable learning outcomes. However, while the Explorables were generally effective, topic-dependent variations emerged. The LLM topic showed declines, possibly due to the probabilistic and less intuitive nature of language models—unlike Diversity and Fairness, which are grounded in more concrete ethical frameworks. These findings suggest that certain AI topics, such as generative models, may require additional scaffolding—such as guided examples or case-based reasoning—to fully support conceptual understanding.

\subsection{Engagement Analysis and Learning Outcomes}
Our analysis of engagement patterns provides key insights into how learners interact with interactive AI education tools. Scrolling behaviors varied widely, reflecting distinct information processing strategies, from linear to exploratory. Surprisingly, no uniformly significant correlation emerged between raw interaction frequency (e.g., scrolling depth, number of interactions) and learning outcomes. This suggests that the mere quantity of engagement does not necessarily translate to better understanding. Instead, we hypothesize that the quality of interaction—moments of reflection, problem-solving, and revisiting key sections—may be more critical. These findings call for more sophisticated engagement metrics and suggest that dynamic scaffolding, especially in later sections, may help sustain attention. Future AI education tools should adopt adaptive interfaces to guide more effective engagement.



\subsection{Design Considerations for Future AI Literacy Tools}
The effectiveness of interactive simulations underscores the need for AI education tools that encourage exploration and reflection. However, our findings also suggest that not all interactive experiences are equally beneficial—some AI topics may require additional scaffolding to support comprehension. Importantly, our regression models show that prior knowledge—captured through pre-test scores—is a strong predictor of learning outcomes, indicating that instructional tools could benefit from adapting to learners’ baseline understanding. While raw engagement metrics (e.g., scrolling) were not predictive of performance, future platforms may be improved by identifying and supporting high-quality engagement behaviors. AI education tools could incorporate real-time feedback to adjust difficulty levels, recommend targeted resources, or prompt users to revisit key concepts. Another important design consideration is balancing interactivity with clarity: more engagement alone does not guarantee learning. Instead, effective tools should foster meaningful engagement through prompts, scaffolding, or contextual explanations.

\subsection{Limitations and Threats to Validity}
While this study offers valuable insights, several limitations should be noted. Our self-selected online sample may introduce biases in engagement and prior AI knowledge, underscoring the need for broader participant diversity. Additionally, reliance on quantitative engagement metrics, such as scrolling behavior, provides an incomplete view of cognitive engagement; future research should incorporate qualitative methods like think-aloud protocols. Finally, our study focused on short-term learning gains, leaving long-term retention and knowledge transfer unexamined. Future work should explore sustained learning effects and the applicability of AI literacy skills in new contexts.

\section{Conclusion}

This study provides compelling evidence for the effectiveness of interactive simulations in advancing AI literacy, particularly in fostering conceptual engagement, confidence, and generalization across topics. These simulations foster deeper conceptual engagement and critical analysis by enabling learners to ``think like a scientist'' through hypothesis testing, experimentation, and real-time observation of AI behavior. However, our findings also reveal that basic engagement metrics do not reliably predict learning outcomes. This underscores the importance of designing AI education tools prioritizing meaningful interaction over surface-level interactivity. As AI systems increasingly shape societal decision-making, these insights highlight the need for thoughtfully designed educational interventions that equip individuals with the skills to assess and navigate AI-driven technologies critically.

\section{Acknowledgments}
We thank the reviewers for the comments and feedback. This work was supported through a seed grant from Stanford Accelerator for Learning.



%
%
%
\clearpage
\bibliographystyle{splncs04}
\bibliography{99_refs}

\begin{thebibliography}{10}
\providecommand{\url}[1]{\texttt{#1}}
\providecommand{\urlprefix}{URL }
\providecommand{\doi}[1]{https://doi.org/#1}

\bibitem{alam2022digital}
Alam, A.: A digital game based learning approach for effective curriculum transaction for teaching-learning of artificial intelligence and machine learning. In: 2022 International Conference on Sustainable Computing and Data Communication Systems (ICSCDS). pp. 69--74. IEEE (2022)

\bibitem{anik2021data}
Anik, A.I., Bunt, A.: Data-centric explanations: explaining training data of machine learning systems to promote transparency. In: Proceedings of the 2021 CHI Conference on Human Factors in Computing Systems. pp. 1--13 (2021)

\bibitem{carney2020teachable}
Carney, M., Webster, B., Alvarado, I., Phillips, K., Howell, N., Griffith, J., Jongejan, J., Pitaru, A., Chen, A.: Teachable machine: Approachable web-based tool for exploring machine learning classification. In: Extended abstracts of the 2020 CHI conference on human factors in computing systems. pp.~1--8 (2020)

\bibitem{carolus2023mails}
Carolus, A., Koch, M., Straka, S., Latoschik, M.E., Wienrich, C.: Mails -- meta {AI} literacy scale: Development and testing of an {AI} literacy questionnaire based on well-founded competency models and psychological change- and meta-competencies (2023)

\bibitem{casal2023ai}
Casal-Otero, L., Catala, A., Fern{\'a}ndez-Morante, C., Taboada, M., Cebreiro, B., Barro, S.: Ai literacy in k-12: a systematic literature review. International Journal of STEM Education  \textbf{10}(1), ~29 (2023)

\bibitem{chi2014icap}
Chi, M.T., Wylie, R.: The icap framework: Linking cognitive engagement to active learning outcomes. Educational psychologist  \textbf{49}(4),  219--243 (2014)

\bibitem{de1998scientific}
De~Jong, T., Van~Joolingen, W.R.: Scientific discovery learning with computer simulations of conceptual domains. Review of educational research  \textbf{68}(2),  179--201 (1998)

\bibitem{de2024human}
De~Silva, D., Jayatilleke, S., El-Ayoubi, M., Issadeen, Z., Moraliyage, H., Mills, N.: The human-centred design of a universal module for artificial intelligence literacy in tertiary education institutions. Machine Learning and Knowledge Extraction  \textbf{6}(2),  1114--1125 (2024)

\bibitem{dignum2019responsible}
Dignum, V.: Responsible artificial intelligence: how to develop and use AI in a responsible way, vol.~2156. Springer (2019)

\bibitem{goeser2018view}
Goeser, P.T., Hamza-Lup, F.G., Johnson, W.M., Scharfer, D.: View: A virtual interactive web-based learning environment for engineering (2018)

\bibitem{googlePAIRExplorables}
Google: Explorables - google pair. \url{https://pair.withgoogle.com/explorables/} (2025), accessed: 2025-01-22

\bibitem{griffin2014assessment}
Griffin, P., Care, E.: Assessment and teaching of 21st century skills: Methods and approach. Springer (2014)

\bibitem{hallgren2002interactive}
Hallgren, R.C., Parkhurst, P.E., Monson, C.L., Crewe, N.M.: An interactive, web-based tool for learning anatomic landmarks. Academic Medicine  \textbf{77}(3),  263--265 (2002)

\bibitem{hidalgo2024agency}
Hidalgo, C.: Agency in ai and education policy: European resolution three on harnessing the potential for ai in and through education. In: International Conference on Artificial Intelligence in Education. pp. 319--327. Springer (2024)

\bibitem{hitron2019can}
Hitron, T., Orlev, Y., Wald, I., Shamir, A., Erel, H., Zuckerman, O.: Can children understand machine learning concepts? the effect of uncovering black boxes. In: Proceedings of the 2019 CHI conference on human factors in computing systems. pp. 1--11 (2019)

\bibitem{jatzlau2019s}
Jatzlau, S., Michaeli, T., Seegerer, S., Romeike, R.: It’s not magic after all--machine learning in snap! using reinforcement learning. In: 2019 IEEE blocks and beyond workshop (B\&B). pp. 37--41. IEEE (2019)

\bibitem{kasinidou2021educating}
Kasinidou, M., Kleanthous, S., Orphanou, K., Otterbacher, J.: Educating computer science students about algorithmic fairness, accountability, transparency and ethics. In: Proceedings of the 26th ACM Conference on Innovation and Technology in Computer Science Education V. 1. pp. 484--490 (2021)

\bibitem{kaspersen2022high}
Kaspersen, M.H., Bilstrup, K.E.K., Van~Mechelen, M., Hjort, A., Bouvin, N.O., Petersen, M.G.: High school students exploring machine learning and its societal implications: Opportunities and challenges. International Journal of Child-Computer Interaction  \textbf{34},  100539 (2022)

\bibitem{kim2023examining}
Kim, H.J., So, H.J., Suh, Y.J.: Examining the impact of flipped learning for developing young job seekers’ ai literacy. In: International Conference on Artificial Intelligence in Education. pp. 817--823. Springer (2023)

\bibitem{long2020ai}
Long, D., Magerko, B.: What is ai literacy? competencies and design considerations. In: Proceedings of the 2020 CHI conference on human factors in computing systems. pp. 1--16 (2020)

\bibitem{mioduser2000web}
Mioduser, D., Nachmias, R., Lahav, O., Oren, A.: Web-based learning environments: Current pedagogical and technological state. Journal of research on computing in education  \textbf{33}(1),  55--76 (2000)

\bibitem{newman2025ai}
Newman-Griffis, D.: Ai thinking: a framework for rethinking artificial intelligence in practice. Royal Society Open Science  \textbf{12}(1),  241482 (2025)

\bibitem{ng2021conceptualizing}
Ng, D.T.K., Leung, J.K.L., Chu, S.K.W., Qiao, M.S.: Conceptualizing ai literacy: An exploratory review. Computers and Education: Artificial Intelligence  \textbf{2},  100041 (2021)

\bibitem{parker2001interactive}
PARKER, M.J., SEIFTER, J.L.: An interactive, web-based learning environment for pathophysiology. Academic Medicine  \textbf{76}(5), ~550 (2001)

\bibitem{popa2024critical}
Popa, D.M.: Critical exploratory investigation of ai consumption, ai perception and ai literacy requirements. AI perception and AI literacy requirements (November 01, 2024)  (2024)

\bibitem{prolific}
Prolific: Prolific - participant recruitment for online research. \url{https://www.prolific.com/} (2025), accessed: 2025-01-22

\bibitem{raub2018bots}
Raub, M.: Bots, bias and big data: artificial intelligence, algorithmic bias and disparate impact liability in hiring practices. Ark. L. Rev.  \textbf{71}, ~529 (2018)

\bibitem{servais2006teaching}
Servais, E.L., LaMorte, W.W., Agarwal, S., Moschetti, W., Mallipattu, S.K., Moulton, S.L.: Teaching surgical decision-making: an interactive, web-based approach. Journal of surgical research  \textbf{134}(1),  102--106 (2006)

\bibitem{shved2024teaching}
Shved, E., Bumbacher, E., Mejia-Domenzain, P., Kapur, M., K{\"a}ser, T.: Teaching and measuring multidimensional inquiry skills using interactive simulations. In: International Conference on Artificial Intelligence in Education. pp. 482--496. Springer (2024)

\bibitem{sneirson2024learning}
Sneirson, M., Chai, J., Howley, I.: A learning approach for increasing ai literacy via xai in informal settings. In: International Conference on Artificial Intelligence in Education. pp. 336--343. Springer (2024)

\bibitem{wan2020smileycluster}
Wan, X., Zhou, X., Ye, Z., Mortensen, C.K., Bai, Z.: Smileycluster: supporting accessible machine learning in k-12 scientific discovery. In: Proceedings of the interaction design and children conference. pp. 23--35 (2020)

\end{thebibliography}

\clearpage
\appendix
\section{Participant Demographics}\label{appendix-demographics}
\begin{figure}[h!]
    \centering
    \includegraphics[width=0.9\textwidth]{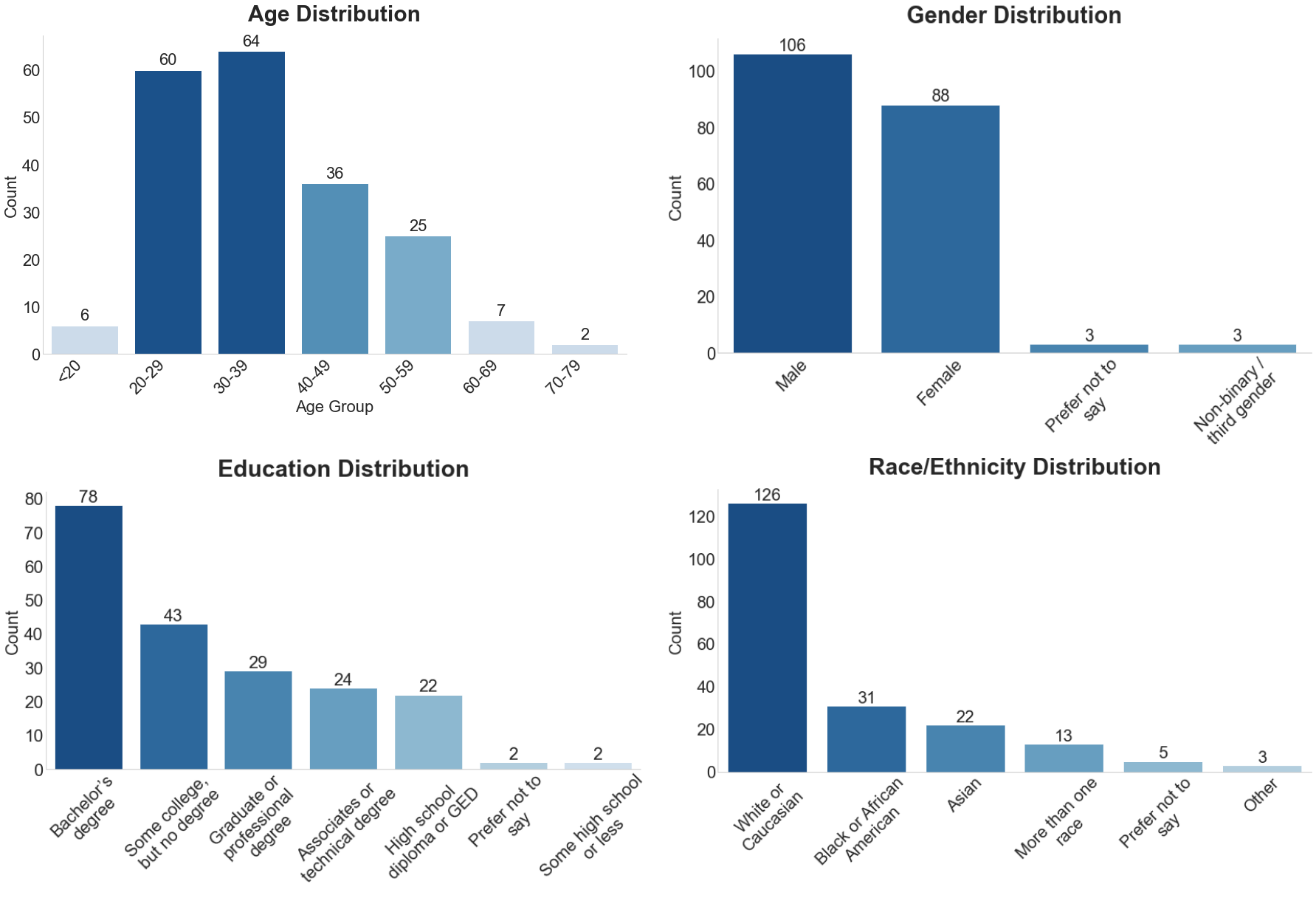}
    \caption{Demographic Distributions of Participants in the Explorable Condition}
    \label{fig:demographics}
\end{figure}

The study encompassed a diverse group of participants, as illustrated in Figure \ref{fig:demographics}. The age distribution showed a right-skewed pattern, with the majority of participants under 40 years old. Gender composition revealed a slight male majority (53\% male, 44\% female, 3\% other). Education levels skewed towards higher education, with bachelor's degree holders being the most common. Racially, White/Caucasian participants formed the largest group, followed by Black or African American and Asian participants.

Notably, a similar distribution was observed in the control groups, ensuring consistency across the study's experimental conditions and enhancing the generalizability of findings across different demographic groups.

\clearpage
\section{Clustering Results}
\begin{figure}[h!]
    \centering
    \includegraphics[width=1\textwidth]{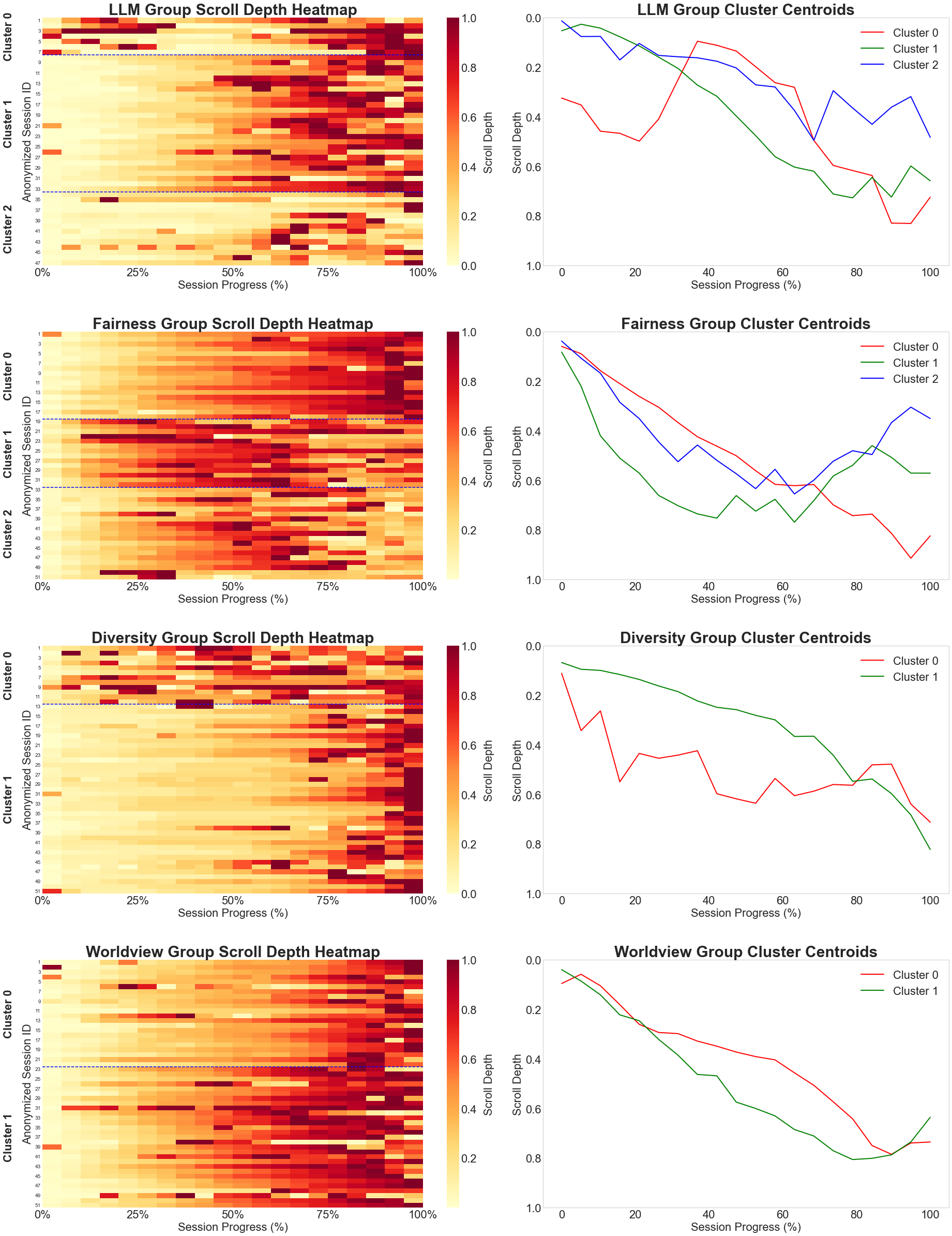}
    \caption{Scroll Pattern Heatmaps and Cluster Centroids}
    \label{fig:heatmap-centroids}
\end{figure}
The analysis revealed varied engagement patterns across AI literacy topics. The LLM and Fairness groups each exhibited three distinct clusters, suggesting more complex content leading to diverse interaction styles. For LLM, patterns included intense focus on early and middle sections, consistent engagement throughout, and initial skimming followed by deep engagement towards the end. Fairness showed moderate engagement with increased mid-section attention, quick early scrolling with deep latter-half engagement, and consistent engagement throughout.

Diversity and Worldview groups displayed two main clusters each, indicating more consistent user approaches. Diversity patterns included variable engagement with alternating deep reading and skimming, and steady, moderate engagement. Worldview showed moderate early engagement increasing towards the end, and quick early scrolling followed by deeper latter-half engagement.

A common trend across all groups was increased engagement towards the latter parts of the content, possibly indicating heightened interest or challenge in these sections. The variety of engagement patterns suggests users approach AI literacy content differently, potentially reflecting diverse learning styles or prior knowledge levels. Some users engage consistently throughout, while others focus on specific sections or gradually increase their engagement over time. These insights could inform the design of more effective, adaptive educational experiences in AI literacy.

\clearpage
\section{Self-reported AI Literacy Questions}
\begin{table}[htbp]
\centering
\begin{tabular}{p{2cm}p{2.5cm}p{7cm}}
\hline
\multicolumn{3}{c}{\textbf{AI Literacy}} \\
\hline
Sources & Item \\
\hline
Use \& Apply AI & Ng et al., 2022 & I can operate AI applications in everyday life. \\
 & & I can use AI applications to make my everyday life easier. \\
 & & I can use artificial intelligence meaningfully to achieve my everyday goals. \\
 & & In everyday life, I can interact with AI in a way that makes my tasks easier. \\
 & & In everyday life, I can work together gainfully with an artificial intelligence. \\
 & & I can communicate gainfully with artificial intelligence in everyday life. \\
\hline
Know \& Understand AI & Ng et al., 2022 & I know the most important concepts of the topic "artificial intelligence". \\
 & & I know definitions of artificial intelligence. \\
 & & I can assess what the limitations and opportunities of using an AI are. \\
 & & I can assess what advantages and disadvantages the use of an artificial intelligence entails. \\
 & & I can think of new uses for AI. \\
 & & I can imagine possible future uses of AI. \\
\hline
Detect AI & Long \& Magerko, 2020 & I can tell if I am dealing with an application based on artificial intelligence. \\
 & Wang et al., 2022 & I can distinguish devices that use AI from devices that do not. \\
 & & I can distinguish if I interact with an AI or a "real human". \\
\hline
AI Ethics & Ng et al., 2022 & I can weigh the consequences of using AI for society. \\
 & & I can incorporate ethical considerations when deciding whether to use data provided by an AI. \\
 & & I can analyze AI-based applications for their ethical implications. \\
\hline
\end{tabular}
\caption{AI Literacy Scale Items and Sources}
\label{table:ai_literacy_scale}
\end{table}

\section{AI Topic Questions}
\subsection{Diversity Group}
\subsubsection{Diversity Group Question Set 1}
\begin{enumerate}
    \item Imagine a leading technology company using an AI system that automatically sifts through resumes to identify the most promising candidates. This system was trained using resumes submitted by all applicants over the past 10 years. Which group of candidates is most likely to be disadvantaged by this algorithm? Select all that apply.
    \begin{itemize}
        \item[$\square$] Candidates with non-traditional educational backgrounds.
        \item[$\square$] Candidates with extensive experience in the tech industry.
        \item[$\square$] Candidates who have recently graduated from top universities.
        \item[$\square$] Candidates from demographic groups historically underrepresented in the tech industry.
    \end{itemize}

    \item Given that the tech industry is predominantly male, and the majority of the resumes used for training came from men, the recruitment system has started to show a preference for male candidates over female candidates. Considering this, what is a more equitable approach to using AI in the recruiting process? Select all that apply.
    \begin{itemize}
        \item[$\square$] Program the AI to disregard gender-related information when reviewing resumes.
        \item[$\square$] Train the AI exclusively with resumes from female candidates to counteract the bias.
        \item[$\square$] Create a balanced training dataset that includes a diverse range of genders, backgrounds, and experiences.
        \item[$\square$] Employ the AI to forecast the success of candidates based on the achievements of past employees, without considering diversity and inclusion.
    \end{itemize}
\end{enumerate}

\subsubsection{Diversity Group Question Set 2}
\begin{enumerate}
\item Imagine the National Home Office deploying an algorithm to detect sham marriages, where couples get married to avoid immigration law, rather than because they have a genuine relationship. This algorithm assigns each couple a 'Red' or 'Green' rating. A 'Red' rating suggests that the Home Office should investigate the couple to confirm or dismiss suspicions of sham activity, while a 'Green' rating indicates that an investigation is not warranted. The algorithm was trained using historical information from proposed sham marriages referred to the triage system over an unspecified three-year period. What potential issues do you see with this algorithm? Select all that apply.
    \begin{itemize}
        \item[$\square$] The algorithm could accurately detect all instances of sham marriages without any biases.
        \item[$\square$] The algorithm will enhance its accuracy in detecting sham marriages over time without any further adjustments.
        \item[$\square$] The algorithm may disproportionately flag marriages involving individuals from certain countries or ethnic backgrounds as 'Red'.
        \item[$\square$] The algorithm might overlook new or evolving patterns of sham marriages not represented in the historical data.
    \end{itemize}

    \item What measures could be implemented to refine the algorithm and ensure it more accurately and equitably identifies sham marriages, without perpetuating biases or overlooking evolving patterns? Select all that apply.
    \begin{itemize}
        \item[$\square$] Regularly update the algorithm with new data, including both successful and unsuccessful marriage cases, to capture evolving trends.
        \item[$\square$] Increase the complexity of the algorithm's decision-making process by adding more historical data, assuming that additional data will automatically correct any biases.
        \item[$\square$] Introduce a feedback mechanism where human reviewers can flag inaccuracies in the algorithm's decisions, facilitating continuous learning and improvement.
        \item[$\square$] Rely exclusively on the algorithm's decisions without human oversight, assuming its initial training ensures fairness and accuracy.
    \end{itemize}
\end{enumerate}

\subsection{Fairness Group}
\subsubsection{Fairness Group Question Set 1}
\begin{enumerate}
    \item Imagine a leading technology and online retail company using an algorithm to oversee and evaluate the performance of contract delivery drivers. Drivers are assessed based on a range of variables, including on-time performance, how well the package is hidden from street view, and the driver's ability to fulfill customer requests. Drivers' accounts are terminated if their ratings fall below an acceptable level. Which statement(s) about this algorithm are true? Select all that apply.
    \begin{itemize}
        \item[$\square$] The algorithm might not accurately account for variations in customer feedback quality, where some customers are more critical than others, leading to inconsistent ratings for drivers.
        \item[$\square$] The algorithm drastically improves evaluation efficiency, ensuring an absolutely fair assessment of every driver's performance.
        \item[$\square$] External factors, such as weather conditions or package handling by previous supply chain segments, could unfairly affect a driver's performance rating.
        \item[$\square$] Human supervision is not necessary because this algorithm provides a comprehensive assessment of driver performance.
    \end{itemize}

    \item What steps can be taken to enhance the fairness and accuracy of the algorithmic evaluation system for contract drivers, ensuring it accounts for external factors and provides equitable assessments? Select all that apply.
    \begin{itemize}
        \item[$\square$] The idea of increasing the frequency of evaluations to weekly, assuming that more data points will automatically lead to fairer assessments.
        \item[$\square$] Implement human intervention where drivers can appeal against low ratings, allowing for a review of the circumstances leading to those ratings.
        \item[$\square$] Introduce a mechanism to normalize ratings based on external factors like weather conditions and customer rating patterns, ensuring a more equitable evaluation.
        \item[$\square$] Make the model more aggressive to increase the accuracy of assessment.
    \end{itemize}
\end{enumerate}
\subsubsection{Fairness Group Question Set 2}
\begin{enumerate}
    \item Imagine a police department using a computerized monitoring system to locate gunshots. A network of 22 acoustic listening devices, fine-tuned to distinguish gunshots from other sounds, has been set up to cover neighborhoods with the highest crime rates. Identify the FALSE statement about this scenario.
    \begin{itemize}
        \item[$\square$] Deploying surveillance technology in socioeconomically disadvantaged neighborhoods could inadvertently reinforce biases and exacerbate issues like over-policing and racial profiling.
        \item[$\square$] The effectiveness and accuracy of acoustic listening devices can vary across different environments, potentially leading to disparities in security resource allocation and community policing.
        \item[$\square$] The system's deployment provides the police with a dependable and accurate means of locating gunshots, thus eliminating the effort required for evidence collection and investigation.
        \item[$\square$] Transparency and accountability in the operation of these systems and the decision-making process based on their alerts are crucial for maintaining community trust and ensuring ethical policing practices.
    \end{itemize}

    \item Considering the concerns around the deployment of acoustic listening devices to detect gunshots in certain neighborhoods, what steps could be taken to address the fairness and effectiveness issues identified? Select all that apply.
    \begin{itemize}
        \item[$\square$] Focus solely on the technological aspects to improve the system's accuracy and efficiency.
        \item[$\square$] Eliminate the use of acoustic listening devices altogether, relying on traditional policing methods without technological assistance.
        \item[$\square$] Engage with community leaders and residents from the neighborhoods under surveillance to gather feedback and adjust the deployment strategy accordingly.
        \item[$\square$] Ensure the acoustic listening devices are equally distributed across all communities.
    \end{itemize}
\end{enumerate}

\subsection{LLM Group}
\subsubsection{LLM Group Question Set 1}
\begin{enumerate}
    \item Imagine a lawyer employing a large language model like ChatGPT to draft a legal brief for court, wherein the model cites previous cases. What potential issues could arise from this scenario? Select all that apply.
    \begin{itemize}
        \item[$\square$] AI-Enhanced Persuasion: The AI's use of language is so persuasive it unfairly influences court decisions.
        \item[$\square$] The model might provide incorrect or outdated information regarding case law.
        \item[$\square$] The AI's training data could introduce or perpetuate biases in legal reasoning or case selection.
        \item[$\square$] AI generates new legal theories and insights previously unconsidered by human lawyers.
    \end{itemize}

    \item Considering the scenario where a lawyer might use a large language model (LLM) like ChatGPT to draft a legal brief for court, which includes citing non-existing cases, how could LLMs be more effectively and appropriately utilized in this context? Select all that apply.
    \begin{itemize}
        \item[$\square$] Use LLMs for initial case law research to identify relevant precedents before thorough verification by legal professionals.
        \item[$\square$] Implement LLMs in courtrooms to provide real-time legal advice and objections during trial proceedings.
        \item[$\square$] Employ LLMs to generate ideas for legal arguments and strategies that lawyers can further develop and refine.
        \item[$\square$] Utilize LLMs to draft sections of the legal brief, while ensuring thorough verification by legal professionals, thereby saving time for more complex and nuanced legal work.
    \end{itemize}
\end{enumerate}

\subsubsection{LLM Group Question Set 2}
\begin{enumerate}
    \item Imagine a leading technology company has developed a large language model designed to generate summaries of scientific literature. This model was trained on a vast and carefully curated corpus of humanity's scientific knowledge, comprising over 48 million papers, textbooks, lecture notes, scientific websites, and encyclopedias. Despite its promising capabilities, the company decommissioned the model just three days after its launch. What potential issues might arise from this scientific summary generator? Select all that apply.
    \begin{itemize}
        \item[$\square$] The model could inadvertently produce and disseminate inaccurate or misleading scientific information, undermining trust in scientific literature.
        \item[$\square$] The model was capable of replacing human researchers entirely, making the scientific community redundant.
        \item[$\square$] There might be concerns about intellectual property rights and plagiarism, as the model could generate content that closely resembles existing copyrighted material without proper attribution.
        \item[$\square$] There's a risk of the model reinforcing existing biases found in the training data, potentially skewing the representation of research topics and perspectives in the generated literature.
    \end{itemize}

    \item Given the scenario of the decommissioned large language model designed for generating scientific literature, how could large language models be more effectively and responsibly used to assist in scientific research? Select all that apply.
    \begin{itemize}
        \item[$\square$] Implement stringent quality control measures, including expert review processes, to verify the accuracy and relevance of the content generated by the model before publication.
        \item[$\square$] Depend on the model to make critical decisions in experimental design and interpretation of results, minimizing human effort in research processes.
        \item[$\square$] Use the model to draft initial research ideas and summaries that can then be expanded upon and validated by human researchers, ensuring a collaborative approach between AI and scientists.
        \item[$\square$] Develop a feedback system where the scientific community can report inaccuracies or issues with the content generated, facilitating continuous improvement of the model.
    \end{itemize}
\end{enumerate}
\subsection{Worldview Group}
\subsubsection{Worldview Group Question Set 1}
\begin{enumerate}

    \item Imagine the Transportation Security Administration (TSA) uses an image-processing body scanner at airport checkpoints, to detect potential threats that are not necessarily metal. Before a person steps into the full-body scanner, a TSA officer must register the person's gender by pressing a pink button for female or a blue button for male. The body scanner is programmed to trigger alarms when detecting unexpected anatomical features based on the selected gender. Which populations do you anticipate might face issues with this system? Select all that apply.
    \begin{itemize}
        \item[$\square$] Transgender individuals
        \item[$\square$] People with physical disabilities
        \item[$\square$] People with religious head coverings
        \item[$\square$] Pregnant individuals
        \item[$\square$] Individuals with gynecomastia (enlarged male breast tissue)
    \end{itemize}

    \item Considering the limitations and challenges posed by the current binary gender selection system (pink button for female, blue button for male) used in TSA's image-processing body scanners at airport checkpoints, what alternative labeling or identification method would be more respectful and accurate for transgender individuals, as well as for all passengers, to ensure both security and dignity are upheld? Select all that apply.
    \begin{itemize}
        \item[$\square$] Automate the gender identification process using facial recognition technology to determine a passenger's gender, thereby eliminating the need for manual selection by TSA agents.
        \item[$\square$] Introduce an 'Other' or 'Non-Binary' option alongside the traditional male and female choices.
        \item[$\square$] Require all passengers to verbally disclose their gender identity to TSA agents, ensuring the correct button is pressed without considering privacy or dignity.
        \item[$\square$] Use more advanced AI technology to detect security threats without requiring manual gender selection, focusing on improving the accuracy of analyzing non-metal objects.
    \end{itemize}
\end{enumerate}
\subsubsection{Worldview Group Question Set 2}
\begin{enumerate}
    \item Imagine a prestigious university using an algorithm to assess applicants for a PhD in the respected computer science department. The software was trained using details from previously accepted students, aiming to teach the system to identify candidates the school would likely favor and highlight them to staff, who would make the final call on the applications. What potential issues do you see with this system? Select all that apply.
    \begin{itemize}
        \item[$\square$] The algorithm will not be good at detecting skills related to computer science abilities.
        \item[$\square$] The algorithm may perpetuate biases present in the data of previously accepted students, leading to unfair treatment of applicants from underrepresented backgrounds.
        \item[$\square$] There could be difficulties in understanding or explaining the algorithm's decision-making process.
        \item[$\square$] This system will increase the workload for the admissions committee to review and select applicants.
    \end{itemize}

    \item Given the potential issues identified with using a machine learning algorithm for assessing PhD applicants based on data from previously accepted students, what would be a better approach for the university to take to address these concerns while still leveraging technology to assist in the admissions process? Select all that apply.
    \begin{itemize}
        \item[$\square$] Train the model to rely solely on quantitative metrics such as GPA and standardized test scores when selecting top candidates to avoid potential biases related to diversity.
        \item[$\square$] Combine algorithmic recommendations with thorough human review, ensuring that the final decision benefits from both AI insights and human judgment.
        \item[$\square$] Implement a system where the algorithm's decisions are periodically reviewed only when applicants challenge their rejection, to minimize the workload on the admissions committee.
        \item[$\square$] Regularly audit the algorithm for biases and implement bias correction techniques to adjust the decision-making process.
    \end{itemize}
\end{enumerate}

\clearpage
\subsection{Question Selection for Explorable Group}
\begin{table}[htbp]
\centering
\resizebox{\textwidth}{!}{%
\begin{tabular}{>{\centering\arraybackslash}p{0.15\textwidth}>{\centering\arraybackslash}p{0.2\textwidth}>{\centering\arraybackslash}p{0.55\textwidth}}
\toprule
\textbf{Explorable Group} & \textbf{Pre-test Question} & \textbf{Post-test Questions} \\
\midrule
LLM & LLM Q2 & LLM Q1, Diversity Q1, Fairness Q1, Worldview Q1 \\
\addlinespace
Worldviews & Worldview Q1 & Worldview Q2, Fairness Q2, Diversity Q1, LLM Q2 \\
\addlinespace
Diversity & Diversity Q1 & Diversity Q2, Worldview Q2, Fairness Q2, LLM Q1 \\
\addlinespace
Fairness & Fairness Q2 & Fairness Q1, Worldview Q1, Diversity Q2, LLM Q2 \\
\bottomrule
\end{tabular}%
}
\caption{Question Selection for Pre-test and Post-test Surveys}
\label{table:question_selection}
\end{table}

\section{Pre-Test Questionnaire}
\begin{enumerate}
    \item How old are you?

    \item How do you describe yourself?
    \begin{itemize}
        \item[$\square$] Male
        \item[$\square$] Female
        \item[$\square$] Non-binary / third gender
        \item[$\square$] Prefer to self-describe
        \item[$\square$] Prefer not to say
    \end{itemize}

    \item In which country do you currently reside?

    \item What is the highest level of education you have completed?
    \begin{itemize}
        \item[$\square$] Some high school or less
        \item[$\square$] High school diploma or GED
        \item[$\square$] Some college, but no degree
        \item[$\square$] Associates or technical degree
        \item[$\square$] Bachelor's degree
        \item[$\square$] Graduate or professional degree (MA, MS, MBA, PhD, JD, MD, DDS etc.)
        \item[$\square$] Prefer not to say
    \end{itemize}

    \item Choose one or more races that you consider yourself to be
    \begin{itemize}
        \item[$\square$] White or Caucasian
        \item[$\square$] Black or African American
        \item[$\square$] American Indian/Native American or Alaska Native
        \item[$\square$] Asian
        \item[$\square$] Native Hawaiian or Other Pacific Islander
        \item[$\square$] Other
        \item[$\square$] Prefer not to say
    \end{itemize}

    \item To what extent do you agree or disagree with the following statements: AI Literacy Scale (\ref{table:ai_literacy_scale})

    \item Explorable Group Target AI Question Set
\end{enumerate}

\section{Post-Test Questionnaire}
\begin{enumerate}

    \item Explorable Group Target AI Question Set
    \item Non-target AI Question Set 1
    \item Non-target AI Question Set 2
    \item Non-target AI Question Set 3
    \item To ensure you're paying attention, please select "Reinforcement Learning" from the list below.
    \begin{itemize}
        \item[$\bigcirc$] Supervised Learning
        \item[$\bigcirc$] Unsupervised Learning
        \item[$\bigcirc$] Reinforcement Learning
        \item[$\bigcirc$] Transfer Learning
    \end{itemize}
    \textit{Note: The order of these question sets (items 1-5) was randomized for each participant.}
    
    \item To what extent do you agree or disagree with the following statements: AI Literacy Scale (\ref{table:ai_literacy_scale})

    \item What other questions you might have but were not answered by the explorable explanation?
\end{enumerate}

\clearpage
\section{Tables}

\begin{table*}[h]
\centering
\caption{Performance Metrics Across Conditions}
\label{tab:performance_metrics_table}
\small
\begin{tabular}{@{}llcccccccccc@{}}
\toprule
\multirow{2}{*}{Condition} & \multirow{2}{*}{AI Topic} & \multicolumn{3}{c}{Identify Issue} & \multicolumn{3}{c}{Propose Solutions} & \multicolumn{3}{c}{Overall} \\
\cmidrule(lr){3-5} \cmidrule(lr){6-8} \cmidrule(l){9-11}
& & Pre & Post & P-value & Pre & Post & P-value & Pre & Post & P-value \\
\midrule
\multirow{4}{*}{Explorable}
& Worldview & 0.46 & 0.65 & \textcolor{darkgreen}{\textbf{0.01}} & 0.60 & 0.69 & 0.18 & 0.53 & 0.67 & \textcolor{darkgreen}{\textbf{0.00}} \\
& Diversity & 0.74 & 0.76 & 0.73 & 0.65 & 0.86 & \textcolor{darkgreen}{\textbf{0.00}} & 0.69 & 0.81 & \textcolor{darkgreen}{\textbf{0.00}} \\
& Fairness & 0.32 & 0.77 & \textcolor{darkgreen}{\textbf{0.00}} & 0.61 & 0.78 & \textcolor{darkgreen}{\textbf{0.01}} & 0.47 & 0.77 & \textcolor{darkgreen}{\textbf{0.00}} \\
& LLM & 0.77 & 0.68 & 0.13 & 0.76 & 0.63 & \textcolor{darkgreen}{\textbf{0.01}} & 0.76 & 0.66 & \textcolor{darkgreen}{\textbf{0.01}} \\
\midrule
\multirow{4}{*}{Basic Control}
& Worldview & 0.46 & 0.64 & \textcolor{darkgreen}{\textbf{0.02}} & 0.46 & 0.52 & 0.48 & 0.46 & 0.58 & \textcolor{darkgreen}{\textbf{0.05}} \\
& Diversity & 0.57 & 0.71 & 0.05 & 0.69 & 0.72 & 0.65 & 0.63 & 0.71 & 0.06 \\
& Fairness & 0.25 & 0.69 & \textcolor{darkgreen}{\textbf{0.00}} & 0.54 & 0.80 & \textcolor{darkgreen}{\textbf{0.00}} & 0.39 & 0.75 & \textcolor{darkgreen}{\textbf{0.00}} \\
& LLM & 0.70 & 0.65 & 0.33 & 0.68 & 0.61 & 0.15 & 0.69 & 0.63 & 0.11 \\
\midrule
\multirow{4}{*}{Static Control} 
& Worldview & 0.37 & 0.70 & \textcolor{darkgreen}{\textbf{0.00}} & 0.47 & 0.66 & \textcolor{darkgreen}{\textbf{0.01}} & 0.42 & 0.68 & \textcolor{darkgreen}{\textbf{0.00}} \\
& Diversity & 0.58 & 0.74 & \textcolor{darkgreen}{\textbf{0.03}} & 0.62 & 0.74 & \textcolor{darkgreen}{\textbf{0.05}} & 0.60 & 0.74 & \textcolor{darkgreen}{\textbf{0.01}} \\
& Fairness & 0.21 & 0.75 & \textcolor{darkgreen}{\textbf{0.00}} & 0.61 & 0.65 & 0.57 & 0.41 & 0.70 & \textcolor{darkgreen}{\textbf{0.00}} \\
& LLM & 0.77 & 0.68 & 0.15 & 0.81 & 0.68 & \textcolor{darkgreen}{\textbf{0.01}} & 0.79 & 0.68 & \textcolor{darkgreen}{\textbf{0.01}} \\
\bottomrule
\end{tabular}
\caption*{\textit{Note: \textcolor{darkgreen}{\textbf{Bold dark}} values indicate significant results ($p \leq 0.05$), suggesting significant changes in performance.}}
\end{table*}
\end{document}